\documentclass[10pt,journal]{IEEEtran}
\newtheorem{theorem}{Theorem}

\newtheorem{lemma}{Lemma}

\newtheorem{remark}{Remark}

\newtheorem{assumption}{Assumption}
\newtheorem{corollary}{Corollary}
\newtheorem{proposition}{Proposition}
\usepackage{graphicx}
\usepackage{subfigure}
\usepackage{array}
\usepackage{amsmath}
\usepackage{amssymb}
\usepackage{cite}
\usepackage{enumerate} 
\usepackage{float}
\usepackage{amsfonts} 
\usepackage{epsfig} 
\usepackage{epstopdf}
\usepackage[dvipsnames]{xcolor}
\usepackage{tikz}\newcommand*{\circled}[1]{\lower.7ex\hbox{\tikz\draw (0pt, 0pt)%
 		circle (.4em) node {\makebox[1em][c]{\small #1}};}}
\usepackage{color}
\usepackage[scaled]{helvet}

\begin{document}
	\title{Backstepping Design Embedded With Time-Varying Command Filters}
	\author{Hefu Ye,   and Yongduan Song$^*$, \IEEEmembership{Fellow, IEEE}
		\thanks{This work was supported by the National Natural Science Foundation
		of China under grant (No.61991400, No.61991403, No.61860206008, and No.61933012). (Corresponding Author: Yongduan Song.)}	
		\thanks{H. F. Ye   is with Chongqing Key Laboratory of Autonomous Systems, Institute of Artificial Intelligence, School of Automation, Chongqing University, Chongqing 400044, China, and also with Star Institute of Intelligent Systems (SIIS), Chongqing 400044, China.   (e-mail:    yehefu@cqu.edu.cn).}
		 	\thanks{Y. D. Song is with Chongqing Key Laboratory of Autonomous Systems, Institute of Artificial Intelligence, School of Automation, Chongqing University, Chongqing 400044, China. 
		 	(e-mail:   
		 	ydsong@cqu.edu.cn).}
	}

\markboth{IEEE Transactions on Systems, Man, and Cybernetics: Systems, Submitted, Nov. 2021.}%
{Ye} 
	\maketitle
	
	\begin{abstract} 
		 If embedded with command filter properly, the implementation of backstepping design could be dramatically simplified. In this paper, we introduce  a   command filter with time-varying gain and integrate it with backstepping design, resulting in a new set of backstepping control algorithms with  low complexity even for high-order strict-feedback systems. Furthermore, with the aid of ``softening" sign function based compensator,  zero-error output tracking is ensured while  at the same time maintaining prescribed  transient performance. Numerical simulation is carried out to verify the effectiveness and benefits of the proposed method.
	\end{abstract}
	
	
	\section{Introduction}
	\label{sec:introduction}
	\IEEEPARstart{B}{ackstepping} technique has been a dominant and popular control design approach   for strict-feedback systems  during the past decades\cite{Krstic1992}, yet it is noted that the resultant control algorithms from backstepping design are complicated,  making it nontrivial for programming and implementation. This is particularly true when system order increases, because at each step of design, the time derivatives of virtual control signals   $\alpha_i~(i=1,\cdots,n-1)$  are required in the next step,  thereby $u(t)$ requires the first-order time derivative of   $\alpha_{n-1}$, which requires the second-order time derivative of  $\alpha_{n-2}$, and so on, rendering the analytical computation of the repetitive derivative of the virtual control quite involved.   	Several efforts,  such as  dynamic surface control (DSC)\cite{Swaroop2000,GeSS2008,Yip1997,zzr}, command filtered backstepping\cite{Polycarpou2009,Polycarpou2012},  {and neural networks/fuzzy logic systems  approximation\cite{Liu-SCIS,Liu-TNNLS,WangHQ-TCAS,KongJie}} have been made in alleviating such complexity.  
	
	Dynamic surface control  uses a first-order filter to get the differential estimation of virtual control signals without the need for complex analytic computation, making it simpler to implement. The estimation errors  accumulate as the system order increases, which inevitably affects the  control performance. Command filtered backstepping method addresses this issue by  introducing a compensator to reduce the estimation error generated by DSC. Also, the rigorous analysis of the effect of the command filter on closed-loop stability is provided in \cite{Polycarpou2009}.  Those methods   have been  applied to spacecraft rendezvous and docking system\cite{Duan2012},  dynamic positioning system of ships\cite{Krstic2016},   multi-input multi-output constrained nonlinear system\cite{Zhao2019}  and many other practical systems.  
	
	Recently,  a fraction dynamic surface is introduced in \cite{Wang2017} to achieve a finite-time differential estimation. The work by \cite{Shi2018} originally reports a finite-time command filtered backstepping technology by exploiting the finite-time differentiator\cite{Levant2003} to ensure that the tracking error converges into a residual set within finite-time, motivating several other    results\cite{Shi2020,Zhao2021-TNN}.  {In addition, neural networks (NN) and fuzzy logic systems (FLS), as effective tools for approximating unknown nonlinear functions, are also often used in backstepping design to eliminate repeated derivations of the virtual control laws.}  	
	However, asymptotic stability as ensured by the classical backstepping method is no longer achieved with those methods, although they  are algebraically simple.   Hence, solving this ``dilemma" is the major motivation of this paper.

	Different from \cite{Shi2018}, where the fractional power feedback based filter and controller  are used to achieve practical finite time convergence, in this paper, we  exploit	a time-varying feedback approach to obtain practical prescribed-time convergence. 
	Actually, it poses significant technical  challenges  when dealing with time-varying filter, time-varying compensator, and time-varying controller synchronously. On one hand, how to introduce the bounded time-varying feedback gain into the design process of filter, compensator and controller synchronously  is an unattempted task. 
	On the other hand, how to ensure that the tracking error can asymptotically converge to zero after the prescribed-time is a meaningful but challenging problem. Most existing works (see \cite{Yu2019-Auto,Zhao2019,Zhao2021-TNN,Shi2018,Shi2020})  can only ensure that the closed-loop signal is ultimately uniformly bounded, rather than asymptotically stable.
	The main contributions of this paper are summarized as follows: 
	\begin{itemize}
		\item  {By integrating command filter with time-varying gains into backstepping technique, the complex algebraic calculations hidden in classical backstepping design is avoided.}
		\item {The filtering errors is reduced by constructing a time-varying error compensation mechanism, and the  closed-loop dynamics of the compensator is guaranteed to be asymptotically stable.}
	\item {In the absence of NN and FLS,  a low algebraic complexity controller is developed for strict-feedback-like systems that is able to achieve zero-error tracking, while maintaining prescribed tracking performance.} 		  
	\end{itemize}

	\section{Problem Formulation}
	Consider the following strict-feedback-like systems\cite{Krishnamurthy 20a} 
	\begin{equation}\label{strict-feedback system}
	\left\{\begin{array}{l}
	\dot{x}_i=g_i(x ,t)x_{i+1}+f_i(x, t),~i=1,\cdots,n-1\\ 
	\dot{x}_n=g_n(x  ,t)u+f_n(x, t) 
	\end{array}\right.
	\end{equation}
	where   $ {x} =\left[x_1,x_2,\cdots,x_n\right]^{\top}\in \mathbb{R}^n$ is the state vector, and $u$ is the control input. It is assumed that the controllability of (\ref{strict-feedback system}) can be guaranteed by reasonable $\{g_i\}_{i=1}^n$ and $\{f_i\}_{i=1}^n$. The control objective is to construct a control law $u$ such that  all the closed-loop signals are bounded and the tracking error $z_1=x_1-x_d$ converges into a disc region  in prescribed-time $T$ and  ultimately converges to zero  as $t\rightarrow+\infty$. 
	
In the subsequent development, the following commonly used assumptions are needed: 
		\begin{assumption}\cite{Polycarpou2009}\cite{Shi2018}
		The  desired trajectory $x_d$ and its derivative $\dot x_d$ are known, bounded and continuous.
	\end{assumption}
	\begin{assumption}\cite{Polycarpou2009}\cite{Shi2018}
		For $i=1,\cdots,n$, the time-varying scalar function  ${f_i(x, t)}$ and ${\dot f_i(x ,t)}$ are known, bounded and continuous. 
	\end{assumption} 
\begin{assumption} \cite{Polycarpou2009}\cite{Shi2018}
	For $i=1,\cdots,n$, the virtual/actual control gain  ${g_i(x ,t)}$ and ${\dot g_i(x ,t)}$ are continuous, known, bounded and away from zero.
\end{assumption}

	\begin{remark}
		System (\ref{strict-feedback system}) belongs to a broader category of strict-feedback system in the mathematical sense, because the virtual/actual control gains $\{g_i\}_{i=1}^n$  and system functions  $\{f_i\}_{i=1}^n$ can involve the ``later" states. Assumption 3 ensures controllability and uniform relative degree. In some existing works studying   strict-feedback or strict-feedback-like systems ({\textit{e.g.,}} \cite{Huang2018,Krishnamurthy 20a} and some references therein), for $i=1,\cdots,n$, the functions   $ {g_i } $ and $ {f_i } $ are assumed to satisfy  $f_i,g_i\in \mathcal{C}^{n-1}$, which is more stringent than Assumptions 2-3 for $n\geq2$. 
	\end{remark} 
	
	\begin{lemma}\label{lemma1} 
		\cite{Huang2018}
		Given any positive smooth function $\sigma(t):[0,+\infty)\rightarrow \mathbb{R}^+$, the following inequality holds
		\begin{equation}
		|s|-\frac{s^2}{\sqrt{s^2+\sigma^2}}<\sigma, ~\forall s\in \mathbb{R}.
		\end{equation} 
	\end{lemma}

	\begin{lemma}\label{lemma2}
		\cite{Song 17} Consider the time-varying function $\mu_0(t)=({T}/({T-t}))^{n+m} $ on $[0,T)$, with positive integers $m,n$. If a continuously differentiable function $V:[0,T)\rightarrow[0,+\infty)$ satisfies
		$	\dot V\leq-k\mu_0(t)V+\mu_0 d^2(t)$ 
		for positive constant $k$ and bounded function $d(t)$, then we have   $V$ is bounded for $[0,T)$. And if $d(t)\equiv0,$ then $\lim_{t\rightarrow T} V=0$.
	\end{lemma}
	
 In order to  defined such time-varying function as in Lemma 2 on $[0,+\infty)$, we reconstruct a bounded  function $\mu(t):[0,+\infty)\rightarrow [1,\bar{\mu}) $, as follows	
	\begin{equation}\label{mu}
\mu(t)=\left\{\begin{array}{l}
\frac{T+\epsilon}{T+\epsilon-t},~~~~~~~~~~~0\leq t< T\\
\bar \mu ,~~~~~~~~~~~~~~~~~t\geq T
\end{array}\right.
\end{equation}
where $0<\epsilon\ll T$ and $\bar\mu\triangleq1+T/\epsilon$. 
	\begin{corollary}\label{corollary1}
		Consider the continuous and bounded function  $\mu(t)$ as defined in (\ref{mu}).  If a continuously differentiable function $V(t):[0,+\infty)\rightarrow[0,+\infty)$ satisfies
		\begin{equation}\label{4}
		\dot V\leq-k\mu (t)V , ~ ~ ~~0\leq t< T
		\end{equation}
		and
		\begin{equation}\label{5}
		\dot V\leq-k\bar\mu V+\bar\mu \sigma(t) ,~ ~ ~~ t\geq T
		\end{equation}
		for positive constant $k$, and the function $\sigma(t)$ satisfying $\lim _{t\rightarrow +\infty}\int_{0}^{t}\sigma (\tau) d\tau\leq \bar\sigma <+\infty$
		with $\bar\sigma $ being a positive constant, then we have that $V (t)$ is bounded and converges to an adjustable set $\Omega=\{V(t)|V(t)\leq e^{-k (T+T^2/\epsilon)}V(0)\}$ as $t\rightarrow T$   and finally converges to zero as $t\rightarrow +\infty$.
	\end{corollary}

	\textit{Proof:} If $0\leq t<T$, according to Lemma \ref{lemma2} we know that $V\in \mathcal{L}_{\infty} [0,T)$. When $t\rightarrow T$, by solving the equation (\ref{4}), we obtain 
	\begin{equation}
	V(t)\leq e^{-k (T+T^2/\epsilon)}V(0).
	\end{equation}
	Thus the trajectory of $V(t)$ converges to the set $\Omega $ as $t\rightarrow T$.
	For $t\geq T$,   integrating the both sides of (\ref{5}) on $[T,t)$, we have 
	\begin{equation}\label{7}
	V(t)\leq V(T)-k\bar\mu\int_{T}^{t}V(\tau) d\tau+\bar\mu\int_{T}^{t}\sigma(\tau) d\tau
	\end{equation}
	which implies that $V(t)$ is bounded, and
	\begin{equation}\label{8}
	\int_{T}^{t}V(\tau) d\tau\leq \frac{V(T)}{k}+\frac{\bar\mu\bar\sigma}{k}\in\mathcal{L}_{\infty}.
	\end{equation}
	From the analysis before, it follows that $V(t)$ is uniformly continuous and that $\lim_{t\rightarrow +\infty}\int_{T}^{t}V(\tau) d \tau$ exists, thus $\lim_{t\rightarrow +\infty}V(t)=0$ by Barbalat's Lemma, which completes the proof. $\hfill\blacksquare$

	\section{Main Results}
	\subsection{Time-varying command   filter}

A new  command   filter with a time-varying gain is designed as
	\begin{equation}\label{command prescribed-time filter}
	\delta_i {\dot{\hat\alpha}}_i=\mu(t)(\alpha_{i}-{\hat{\alpha}}_i),~{\hat\alpha}_i(0)=\alpha_i(0),~i=1,\cdots,n
	\end{equation}
	where $\mu(t)$ is defined in (\ref{mu}), and $\delta_i>0$ is a  constant, $\alpha_i$ is the virtual control, which is used as the input of the filter, and $\hat\alpha_i$ is the output of the filter.

	\begin{remark}
 {Different from  most filter based design methods that exploit  a constant high-gain (see,\cite{Zhao2019, Zhao2021-TNN,Polycarpou2012,Swaroop2000} and some references therein),  we revisit the classical first-order filter by a bounded time-varying filter gain in (\ref{command prescribed-time filter}).} 
Note that $({\hat\alpha}_i-\alpha_{i})$ represents the unachieved portion of the virtual control in $i$-th channel, it will be  used as input signals of the compensator as shown in the sequel. On the contrary, the filter errors (the unachieved portions) caused by the first-order filter are not considered in DSC method, which  sacrifices the desired  control performance.
	\end{remark}

	\subsection{Time-varying error compensator} 
	To proceed, we  use the compensated  signal $\{\zeta_i\}_{i=1}^n$ to attenuate  the effect of the filter error $({\hat\alpha}_i-\alpha_{i})$. 	
	Then,  a new error compensator,  combining bounded time-varying gain and ``softening" sign function, is designed as
	\begin{equation}\label{update law}
	\left\{\begin{array}{l}
	\dot\zeta_1=-k_1\mu\zeta_1+g_1({\hat\alpha}_1-\alpha_1)+g_1\zeta_2-\frac{  l_1  \zeta_1}{\sqrt{\zeta_1^2+\sigma_1^2}}\\
	\dot\zeta_i=-k_i\mu\zeta_i+g_i({\hat\alpha}_i-\alpha_i)+g_i\zeta_{i+1}\\
	~~~~~ ~ -g_{i-1}\zeta_{i-1}-\frac{  l_i  \zeta_i}{\sqrt{\zeta_i^2+\sigma_i^2}},~~i=2,\cdots,n-1\\
	\dot\zeta_n=-k_n\mu\zeta_n-g_{n-1}\zeta_{n-1}-\frac{  l_n  \zeta_n}{\sqrt{\zeta_n^2+\sigma_n^2}}
	\end{array}\right.
	\end{equation}
	with $\{\zeta_i(0)\}_{i=1}^n=0$ and $\{\sigma_i(t)\}_{i=1}^n>0$, which satisfying 
	$
	\lim_{t\rightarrow +\infty}\int_{0}^{t}\sigma_i(\tau) d\tau\leq \bar\sigma_i<+\infty
	$
	where $\bar\sigma_i$ is a positive constant,  {\textit{e.g.,} $\sigma_i= e^{-t}$ and $\bar\sigma=1\geq\int_0^{\infty}\sigma(t)dt$.}

	\subsection{Time-varying feedback based controller} 
	As other existing works do, we first define the following coordinate transformation
	\begin{equation}\label{coordinate transformation}
	z_i=x_i-{\hat\alpha}_{i-1},~i=1,\cdots,n
	\end{equation}
	and ${\hat\alpha}_0=x_d$. We	then define the compensated tracking error signals  as
	\begin{equation}\label{tracking error}
	s_i=z_i-\zeta_i,~i=1,\cdots,n.
	\end{equation}
		\begin{remark}
 It is worth noting that each signal $\zeta_i$ is a filtered version of $({\hat\alpha}_i-\alpha_{i})$ and $\{s_i\}_{i=1}^n$ are the compensated tracking errors because they are obtained by eliminating the filter unachieved portion of the corresponding virtual control. These auxiliary signals are used to design  the final actual control input $u$, where the compensating signals are introduced to counteract the impact of the filter errors, allowing the filter output $\hat{\alpha}_i$ to approximate the filter input $\alpha_i$.
	\end{remark}

Then, we construct the virtual/actual control signals as 
	\begin{equation}\label{control input}
\small{	\left\{\begin{array}{l}
	\alpha_1=\frac{1}{g_1}\left(-k_1\mu z_1+\dot{x}_d-f_1-\frac{ l_1  \zeta_1}{\sqrt{\zeta_1^2+\sigma_1^2}}-\frac{s_1}{\sqrt{s_1^2+\sigma_1^2}}\right) \\
	\alpha_i=\frac{1}{g_2}\left(-k_i\mu z_i+\dot{{\hat\alpha}}_{i-1}-f_i-g_{i-1}z_{i-1} \right.\\
	\left.~~~~~~~~~~~~~~~-\frac{ l_i  \zeta_i}{\sqrt{\zeta_i^2+\sigma_i^2}}-\frac{s_i}{\sqrt{s_i^2+\sigma_i^2}}\right), i=2,\cdots,n \\
	u=\frac{1}{g_n}\left(-k_n\mu z_n+\dot{{\hat\alpha}}_{n-1}-f_n-g_{n-1}z_{n-1}\right.\\
	\left.~~~~~~~~~~~~~~~-\frac{  l_n  \zeta_n}{\sqrt{\zeta_n^2+\sigma_n^2}}-\frac{s_n}{\sqrt{s_n^2+\sigma_n^2}}\right)
	\end{array}\right.}
	\end{equation}
	where $\{k_i\}_{i=1}^n$ are positive design parameters and $\mu,$ $\{z_i\}_{i=1}^{n},$ $\{{\dot{{\hat\alpha}}}_i\}_{i=1}^{n-1}$   are defined in the previous subsection. Note that the controller is quite different from the previous controller as mentioned in \cite{Polycarpou2009} and hence we describe the relationship between various signals in a new figure (as shown in Fig. 1), which describes the construction process of the control input $u(t)$ by using the measurable states $\{x_i(t)\}_{i=1}^n$ and the output signals of the time-varying filter and compensator.

	\begin{figure}[!htbp]\label{fig1}
		\centering
		\includegraphics[height=2.7cm]{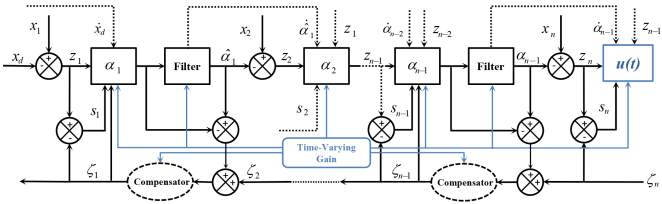}
		\caption{Block diagram of the proposed command filtered backstepping approach.}
	\end{figure}

\begin{remark}	
 {The filter input involves the virtual controller output,  the outputs of the filter and the virtual controller are injected into the compensator dynamics via a subtraction calculation,  meanwhile, the outputs of the compensator and the filter are injected into the controller via  the ``softening" sign function, thereby forming an interconnected closed-loop system. 	
The expressions of (\ref{command prescribed-time filter}), (\ref{update law}) and (\ref{control input}) show the specific structure of the filter, compensator and controller, while revealing  how to deploy the time-varying feedback signal reasonably.  Note that we here do not consider the system in the presence of unmodeled dynamics and parameter uncertainties. The reason for this is that we intend to present a plain scheme that precisely and concisely shows the spirit of the interconnected filter/compensator/controller structural design, instead of exploiting robust scaling to the uncertainties or employing  adaptive laws to estimate   unknown  parameters. This, however, does not mean that the proposed method is not compatible with such robust and/or adaptive schemes.}
\end{remark}
	
	\begin{proposition}\label{proposition1}
		The conclusion that the compensated signal $\zeta_i$ converges to zero as $t$ tends to infinity  holds for any bounded virtual control inputs satisfying $\|{\hat\alpha}_i-\alpha_{i}\|\leq \tau_i$ with $\tau_i$ being positive constant.
	\end{proposition}
	
	\textit{Proof:} Choosing a positive definite  function $V_0=\sum_{i=1}^n \zeta_i^2/2$  as a   Lyapunov function candidate, whose time derivative on $[0,T)$ can be obtained by (\ref{update law}), 
	\begin{equation}\label{dv0}
\small{	\begin{aligned}
	\dot V_0&=-\sum_{i=1}^n\left(k_i\mu\zeta_i^2+\frac{  l_i \zeta_i^2}{\sqrt{\zeta_i^2+\sigma_i^2}}\right)+\sum_{i=1}^{n-1}g_i\zeta_i( {{\hat\alpha}}_i-\alpha_i)
	\end{aligned}}
	\end{equation}
	By using Lemma \ref{lemma1}, one can obtain
	\begin{equation}\label{dv00}
	\begin{aligned}
	\dot V_0&\leq-\sum_{i=1}^n \left(k_i\mu\zeta_i^2-  l_i (\sigma_i-|\zeta_i|)\right) +\sum_{i=1}^{n-1}g_i|\zeta_i|\tau_i\\
	&\leq-\sum_{i=1}^n \left(k_i\mu\zeta_i^2-  l_i \sigma_i\right)  -\sum_{i=1}^{n-1} |\zeta_i|(   l_i-g_i\tau_i)\\
	&\leq-\sum_{i=1}^nk_i\mu\zeta_i^2+ \sum_{i=1}^n  l_i \sigma_i\\
	\end{aligned}
	\end{equation}
	where $l_i$ is chosen such as $l_i\geq  g_i\tau_i$. 
	From Corollary \ref{corollary1}, we have that $\{\zeta_i\}_{i=1}^n$ will converge to zero as $t\rightarrow +\infty.$ $\hfill\blacksquare$

	Now, we are in the position to state the main result.
	\begin{theorem}\label{theorem1}
		Consider system (\ref{strict-feedback system}) under the coordinate transformation (\ref{coordinate transformation}) and the compensated tracking errors (\ref{tracking error}). 
		If we design the command  filter, compensator and the controller as in (\ref{command prescribed-time filter}), (\ref{update law}) and  (\ref{control input}). Then, all  closed-loop signals  are bounded over the entire time domain, and  the tracking error $z_1$ converges into a  computable residual set $ 	\Omega_z $ 
		in prescribed-time $T$ and ultimately decays to zero. 
	\end{theorem}
	
	\textit{Proof:}  	
	\textit{Step 1:} Choose the Lyapunov function candidate as $V_1=0.5s_1^2$, then
	\begin{equation}\label{dv1}
	\small{\begin{aligned}
	\dot V_1&=s_1\dot{s}_1=s_1(\dot{z}_1-\dot\zeta_1)=s_1(\dot x_1-\dot{x}_d-\dot\zeta_1)\\
	&=s_1\left(g_1(z_2+ {{\hat\alpha}}_1)+f_1-\dot{x}_d-\dot\zeta_1\right)\\
	&=s_1\left(g_1\alpha_1+g_1z_2+g_1( {{\hat\alpha}}_1-\alpha_1)+f_1-\dot{x}_d-\dot\zeta_1\right).\\
	\end{aligned}}
	\end{equation}
According to Lemma 2 and inserting  $\dot\zeta_1$ and $\alpha_1$  as defined in (\ref{update law}) and (\ref{control input})   into (\ref{dv1}), yields
	\begin{equation}\label{DV1}
	\begin{aligned}
	\dot V_1=&-k_1\mu s_1^2-\frac{s_1^2}{\sqrt{s_1^2+\sigma_1^2}}+g_1s_1s_2\\
	\leq&-k_1\mu s_1^2-|s_1|+\sigma_1+g_1s_1s_2\\
	\leq&-k_1\mu s_1^2 +\sigma_1+g_1s_1s_2.
	\end{aligned}
	\end{equation}
\textit{Step k:} ($k=2,\cdots,n-1$) Choose the Lyapunov function for the $i$-th subsystem candidate as $V_k=\sum_{i=1}^{k-1}V_i+0.5s_k^2$, whose time derivative is
	\begin{equation}\label{dvk}
	\begin{aligned}
	\dot V_k=&\sum_{i=1}^{k-1}\dot V_i+ s_k\dot{s}_k=\sum_{i=1}^{k-1}\dot V_i+s_k(\dot{z}_k-\dot\zeta_k)\\
	=& \sum_{i=1}^{k-1}\dot V_i-s_k\dot\zeta_k +s_k g_k\alpha_k+\\
	&s_k\left( g_kz_{k+1}+g_k( {{\hat\alpha}}_k-\alpha_k)+f_k-\dot\beta_{k}\right).\\
	\end{aligned}
	\end{equation}
According to the definitions of $\alpha_k$ and $\dot\zeta_k$, we have
	\begin{equation}\label{DVK}
	\begin{aligned}
	\dot V_k=&\sum_{i=1}^k\left(-k_i\mu s_i^2-\frac{s_i^2}{\sqrt{s_i^2+\sigma_i^2}}\right)+g_ks_ks_{k+1}.  
	\end{aligned}
	\end{equation}
	\textit{Step n:} Consider the Lyapunov function candidate as $V_n=\sum_{i=1}^{n-1}V_i+0.5s_n^2$. Similar to the treatment as shown in (\ref{dv1})-(\ref{DV1}), according to the definition of $s_n$,  we have, for $\forall t\in[0,T)$
	\begin{equation}\label{dvn}
 {\small{\begin{aligned}
	\dot V_n=\sum_{i=1}^{n-1}\dot V_i+s_n\dot s_n  
	=&\sum_{i=1}^{n-1}\left(-k_i\mu s_i^2-\frac{s_i^2}{\sqrt{s_i^2+\sigma_i^2}}\right)\\& +g_{n-1}s_{n-1}s_{n}+s_n(\dot z_n-\dot\zeta_n).
	\end{aligned}}}
	\end{equation}
 {It follows from (\ref{update law}), (\ref{tracking error}) and (\ref{control input}) that} 
	\begin{equation}\label{sn}
 {\small{\begin{aligned}
	s_n(\dot z_n-\dot\zeta_n)&=s_n(g_nu+f_n-\dot{{\hat\alpha}}_{n-1}-\dot\zeta_n)\\
	&=s_n\left[\left(-k_n\mu z_n+\dot{{\hat\alpha}}_{n-1}-f_n-g_{n-1}z_{n-1}\right.\right.\\ &\left.\left.~~ -\frac{l_n  \zeta_n}{\sqrt{\zeta_n^2+\sigma_n^2}} -\frac{s_n}{\sqrt{s_n^2+\sigma_n^2}}\right)+f_n-\dot{{\hat\alpha}}_{n-1}\right.\\
	&\left.~~+k_n\mu\zeta_n+g_{n-1}\zeta_{n-1}+\frac{  l_n  \zeta_n}{\sqrt{\zeta_n^2+\sigma_n^2}}\right]\\
	&=-k_n\mu s_n^2-g_{n-1}s_{n-1}s_n-\frac{s_n^2}{\sqrt{s_n^2+\sigma_n^2}}.
	\end{aligned}}}
	\end{equation}
 {Substituting (\ref{sn}) into (\ref{dvn}) yields }
	\begin{equation}\label{dvn1}
 {\begin{aligned}
	\dot V_n =\sum_{i=1}^{n}\left(-k_i\mu s_i^2-\frac{s_i^2}{\sqrt{s_i^2+\sigma_i^2}}\right)\leq -K_1\mu V_n,
	\end{aligned}}
	\end{equation}
	where $K_1=2\min\{k_1,\cdots,k_n\}$. In fact, according to Lemma 1, $\dot V_n$ can also be expressed as, for $t\geq T$
	\begin{equation}\label{dvn2}
	\begin{aligned}
	\dot V_n &\leq\sum_{i=1}^n\left(-k_i\bar\mu s_i^2+\sigma_i-|s_i|\right)\leq-K_1\bar\mu V_n+\Gamma_2, 
	\end{aligned}
	\end{equation}
	where   $\Gamma_2=\sum_{i=1}^n\sigma_i$. From (\ref{dvn1}), (\ref{dvn2}) and Corollary \ref{corollary1}, we have that $V_n$, as well as $\{s_i\}_{i=1}^n$, converge
	to the set 
	\begin{equation}
	\Omega_s=\left\{s_i\Big||s_i|\leq e^{-K(T+T^2/\epsilon)}V_n(0)\right\}
	\end{equation}
	as $t\rightarrow T$ and finally converge to zero as $t\rightarrow +\infty.$ In order to figure out the convergence properties of $\zeta_i$ at $t=T$, we recalling (\ref{dv00}) to get that
	\begin{equation} \label{V_0}
	\begin{aligned}
	\dot V_0 \leq-K_0 V_0+\Gamma_0  
	\end{aligned}
	\end{equation}
	where $K_0=2(1+T/\epsilon)\min\{k_1,\cdots,k_n\}$ and $\Gamma_0=\sum_{i=1}^{n}   l_i \sigma_i(0)$. Integrating both sides of (\ref{V_0}) on $[0,t)$  yields
	$V_0(t)\leq e^{-K_0 t}V_0(0)+ \left(1-e^{-K_0 t}\right){\Gamma_0}/{K_0}$. It is easy to verify that $V_0(t)$ converges to the set
	\begin{equation}
	\Omega_{\zeta} =e^{-K_0 T}V_0(0)+ \left(1-e^{-K_0 T}\right){\Gamma_0}/{K_0}
	\end{equation}
	as $t\rightarrow T$.  Applying Proposition \ref{proposition1} and recalling the fact $z_i=s_i+\zeta_i$, $z_i=x_i- {{\hat\alpha}}_{i-1}$ for $ i=1,\cdots,n $, we know that $\{z_i\}_{i=1}^n$  converge to the set  $
	\Omega_z= \Omega_{\zeta}+\Omega_s $
	as $t\rightarrow T$. Since we have proven that $\lim_{t\rightarrow +\infty}\zeta_i=0$ and $\lim_{t\rightarrow +\infty}s_i=0$, therefore  $\lim_{t\rightarrow +\infty}z_i=0$ can be obtained,   establishing the same for the tracking error $z_1=x_1-x_d$. Furthermore, it is not difficult to verify that all closed-loop signals are continuous and bounded over the entire time domain. $\hfill\blacksquare$
\begin{remark}
 {It should be mentioned that the proposed scheme has the advantages over DSC and command filtered backstepping method [2]-[15]. First, a time-varying error compensation mechanism is introduced to reduce the errors caused by the filter, which, can guarantee that the error compensation signals are practical prescribed-time stable,  reducing the influence of errors timely. Second, different from the UUB or practical finite-time stability achieved in the aforementioned works,  the proposed method guarantees that the tracking error converges to a residual set within prescribed-time and ultimately converges to zero,   a favorable feature in practice.}  
	\end{remark}

	\section{Simulations}
	Consider a electromechanical system   \cite{Shi2018} and rewrite its standard model as the following strict-feedback form:
	\begin{equation}\label{simulation system}
	\left\{\begin{array}{l}
	\dot{x}_1=x_2\\
	\dot{x}_2=x_3-\frac{N}{M}\sin(x_1)-\frac{B}{M}x_2\\
	\dot{x}_3=u-\frac{K_m}{ML}x_2-\frac{H}{ML}x_3\\
	y=x_1.
	\end{array}\right.
	\end{equation}
For simulation, the initial conditions are chosen as $[x_1(0);x_2(0);x_3(0)]=[0.5;0.5;0.5]$, the system parameters are chosen as $M=0.064,~N=3.12,~B=0.02,~K_{m}=0.9,~H=5,~L=15$, and the reference signal is chosen as $x_d=0.5\sin(t)+0.5\sin(0.5t)$. In addition, we set the control parameters   as  $[l_1;l_2;l_3]=[0.1;0.4;20]$, $[k_1;k_2;k_3]=[8;1;1]$, $\epsilon=0.5$, $\delta_1=\delta_2=0.01$, $T=2s$, and $\sigma_1(t)=\sigma_2(t)=\sigma_3(t)=5e^{-0.01t}$. 
Fig. 1 shows that  the tracking process under the proposed method, and Fig. 2 shows  that the inputs and outputs of the prescribed-time command filter and the control input $u$. Fig. 3 shows the trajectories of  the compensator signals $\zeta_i$ and $s_i$. For  comparison, we exploit  the controllers proposed in \cite{Polycarpou2009} and \cite{Shi2018}, the trajectories of the tracking error under different controllers are shown in Fig. 4. These results verify the effectiveness of the proposed method and it can be clearly seen that the  control performance is improved on the basis of existing methods.
	\begin{figure}[!htbp]  
	\centering
	\includegraphics[height=4 cm]{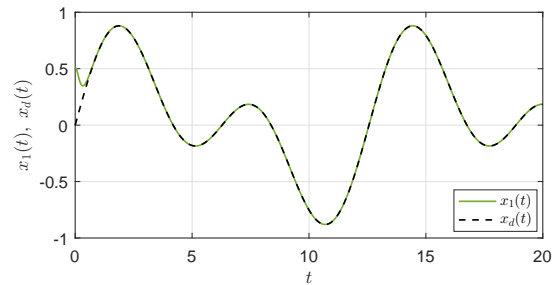}
	\caption{Trajectories of the tracking process.}
\end{figure}

\begin{figure}[!htbp]  
	\centering 
	\includegraphics[height=5.4cm]{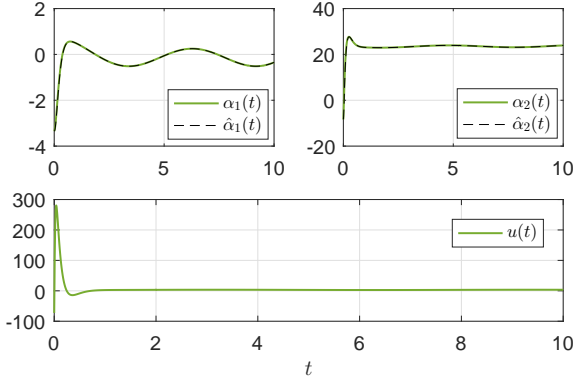} 
	\caption{{Filter output $ {{\hat\alpha}}_1, {{\hat\alpha}}_2$, virtual control input $\alpha_1,\alpha_2$ and actual control input $u$.}}
\end{figure}
\begin{figure}[!htbp]  
	\centering
	\includegraphics[height=4.1 cm]{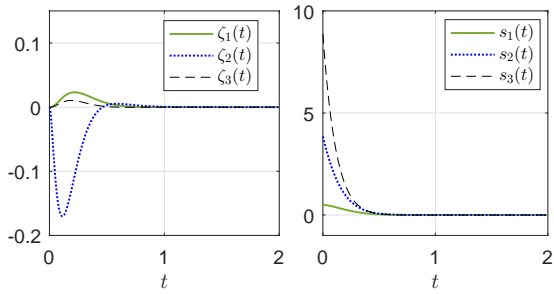} 
	\caption{Trajectories of the compensator signals $\{\zeta_i(t)\}_{i=1}^3$ and the compensated tracking errors $\{s_i(t)\}_{i=1}^3$.}
\end{figure}
\begin{figure}[!htbp]  
	\centering
	\includegraphics[height=4.6cm]{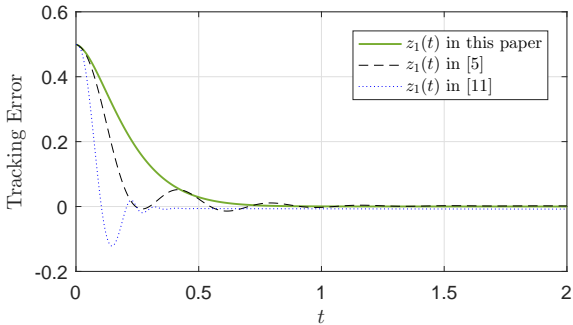} 
	\caption{{Trajectories of the tracking error $z_1$ under different methods.}}
\end{figure} 	
	
		\section{Conclusions} 
	In this work we propose a new method of using command filter to simplicity the implementation of the well known backstepping design technique. By using  the bounded time-varying scaling, we establish the accelerated filter dynamics, and with proper  compensation, we show that the compensator errors converge to zero ultimately and the closed-loop system is asymptotically stable with prescribed performance, opening a new venue for developing backstepping based control algorithms with low-complexity as the issue of ``differential explosion" is completely circumvented.  Future work includes   extending this approach to high-order nonlinear systems with model uncertainties and external disturbances.

\end{document}